\documentclass[runningheads]{llncs}

\usepackage[T1]{fontenc}

\usepackage{algorithm}
\usepackage{algpseudocode}
\usepackage{varwidth}
\usepackage{url}
\usepackage{booktabs}
\usepackage{xcolor} 
\usepackage{listings}
\usepackage{circledsteps}
\usepackage{fontawesome5}
\usepackage{graphicx}
\usepackage{enumitem}

\definecolor{darkgray}{rgb}{.4,.4,.4}
\definecolor{purple}{rgb}{0.65, 0.12, 0.82}
 
\lstdefinelanguage{JavaScript}{
keywords={typeof, new, true, false, catch, function, return, null, catch, switch, var, if, in, while, do, else, case, break},
keywordstyle=\color{blue}\bfseries,
ndkeywords={class, export, boolean, throw, implements, import, this},
ndkeywordstyle=\color{darkgray}\bfseries,
identifierstyle=\color{black},
sensitive=false,
comment=[l]{//},
morecomment=[s]{/*}{*/},
commentstyle=\color{purple}\ttfamily,
stringstyle=\color{red}\ttfamily,
morestring=[b]',
morestring=[b]"
}
 
\newcommand{\node}{Node-RED\;}

\begin{document}

\title{In Specs we Trust? Conformance-Analysis of Implementation to Specifications in Node-RED and Associated Security Risks}

\author{Simon Schneider\inst{1}\orcidID{0000-0001-8605-615X} \and
Komal Kashish\inst{2} \and
Katja Tuma\inst{3}\orcidID{0000-0001-7189-2817} \and 
Riccardo Scandariato\inst{1}\orcidID{0000-0003-3591-7671}}
\authorrunning{S. Schneider et al.}

\institute{Hamburg University of Technology, Hamburg, Germany \and
Siemens Mobility, Berlin, Germany \and
Vrije Universiteit Amsterdam, Amsterdam, The Netherlands}

\titlerunning{In Specs we Trust?}

\maketitle

\begin{abstract}
Low-code development frameworks for IoT platforms offer a simple drag-and-drop mechanism to create applications for the billions of existing IoT devices without the need for extensive programming knowledge.
The security of such software is crucial given the close integration of IoT devices in many highly sensitive areas such as healthcare or home automation.
Node-RED is such a framework, where applications are built from \textit{nodes} that are contributed by open-source developers.
Its reliance on unvetted open-source contributions and lack of security checks raises the concern that the applications could be vulnerable to attacks, thereby imposing a security risk to end users.
The low-code approach suggests, that many users could lack the technical knowledge to mitigate, understand, or even realize such security concerns.

This paper focuses on \textit{hidden} information flows in Node-RED nodes, meaning flows that are not captured by the specifications.
They could (unknowingly or with malicious intent) cause leaks of sensitive information to unauthorized entities.
We report the results of a conformance analysis of all nodes in the Node-RED framework, for which we compared the numbers of specified inputs and outputs of each node against the number of sources and sinks detected with CodeQL.
The results show, that 55\% of all nodes exhibit more possible flows than are specified.
A risk assessment of a subset of the nodes showed, that 28\% of them are associated with a high severity and 36\% with a medium severity rating.


\end{abstract}

\section{Introduction}
\label{sec:intro}

Internet of Things (IoT) devices have seen a tremendous boom in their number over the last decade and have seamlessly become a part of our everyday world. 
From industrial control systems to fitness bands, from smart security systems to medical devices, more and more devices enter the cyberspace continuously to increase the efficiency of their respective industries. 
Predictions for the number of IoT devices being installed, connected, and autonomously managed within the coming years vary, but reach up to 100 billion \cite{huawei}. 

With the increasing adoption of IoT in critical and sensitive domains, the security of such devices becomes also more important, due to the grave financial and personal costs that can occur from compromised systems.
Consequently, research in the field has produced comprehensive studies of the challenges faced (\cite{Ahmad_Alsmadi21_ml_approaches_iot_security,Hassija19_iot_security_survey,Zhang14_Iot_security_challenegs_opportunities}), ongoing work (\cite{Mahmoud15_iot_security,MohamadNoor_Hassan19_iot_security_research}), and possible mitigation approaches (\cite{AlGaradi20_survey_ml_dl_iot,Chatterjee22_iot_anomaly_detection,Williams22_survey_security_iot,Xiao18_iot_security_techniques_ml}).

The increase in demand for physical deployments has also been the propelling factor for rapidly developing IoT software development tools. 
Implementing applications for IoT devices poses some unique challenges for developers~\cite{Udoh18_developing_iot_applications}.
Development frameworks support users in this process and help them realize a wide range of functionality by abstracting many complex activities into simple interfaces, for example, the management of sensor data streams from hardware devices or communication with other entities.
The highest level of abstraction is reached by so-called \textit{no-code} or \textit{low-code} platforms, which simplify the development process into graphical drag-and-drop interfaces, allowing even users with low or no programming knowledge to create applications for IoT devices~\cite{Bock21_low_code_platform,Rokis22_challenges_of_low_code}.

Some development frameworks combine the low-code development with an open-source approach.
Here, the underlying source code that realizes the abstraction of the code-intensive software development into low-code functionalities can be provided by anyone choosing to contribute to the framework.
Moreover, these frameworks not only \textit{allow} the addition of open-source code, they \textit{rely} on it to a large degree.
While this approach leverages the work of numerous open-source developers, and thereby promises to increase the amount of available functionalities, it also introduces security challenges, since the frameworks' code is no longer created by a single authority that can implement secure development practices.
Instead, the responsibility for secure software is shifted to open-source developers.
With insecure software being a major issue for the security of IoT devices~\cite{HaddadPajouh21_security_iot}, this could result in less secure IoT devices.
Although it is possible to enforce security controls on the provided code, this requires large efforts, especially since such controls must be adapted to highly varied code.

Because of the above properties, low-code development frameworks that rely on open-source contributions could potentially provide insufficiently secured software to users that do not have the capabilities to address or even realize this issue.
A possible consequence are vulnerable applications for IoT devices that can be compromised by attackers and cause serious security implications.
The domain of IoT with its intricate connection to many sensitive areas of our lives further increases the seriousness of this.
Additionally to the unaware end user, it could well be that the open-source developers themselves do not realize the extent to which their code could expose sensitive information, given that there are no requirements for security checks or vetting of contributors.
Consequently, the specifications of open-source contributions to the development framework do not accurately depict the reality of the implementation, specifically concerning inputs and outputs over which their functions communicate with other parts of the software.

In this paper, we present the results of a conformance analysis of the nodes in the Node-RED ecosystem, with which we investigated possibly hidden information flows.
In this context, ``hidden'' information flows are flows that can occur in the implemented node but are not considered in the node's specification, i.e., not captured by the specified numbers of inputs and outputs.
These are relevant for the described scenario, since they and their security implications likely go unnoticed by the end user of the ecosystem.
Therefore, we also performed a risk assessment of identified information flows.
The conducted study investigates the above-described security issue faced by low-code open-source development frameworks for IoT devices by shedding light on the extent of information flows that could expose sensitive information to unauthorized entities and are not marked as such by the code's specifications.
Based on the above description, this paper addresses the following research questions:

\vspace{2mm}
\noindent
\fcolorbox{black}{blue!05}{%
    \parbox{0.983\linewidth}{%
        \textbf{\faQuestionCircle\;RQ1: What is the prevalence of hidden information flows in the Node-RED ecosystem?}
    }%
}%

\vspace{2mm}
\noindent 
We performed a conformance analysis of all node packages (the components used to create IoT applications) provided by Node-RED.
The conformance case (convergence, divergence, or absence) of each node package was determined by comparing the number of specified inputs and outputs against the number of sources and sinks detected in an information flow analysis with CodeQL.

\vspace{2mm}
\noindent
\fcolorbox{black}{blue!05}{%
    \parbox{0.983\linewidth}{%
        \textbf{\faQuestionCircle\;RQ2: What are the most likely risks that users of the Node-RED ecosystem are exposed to because of the hidden information flows?}
    }%
}%

\vspace{2mm}
\noindent 
All hidden information flows can be seen as a compliance issue from a software development perspective, but not all hidden information flows necessarily pose a security issue.
To assess the extent to which the detected non-conformances impose security threats to users of the ecosystem, we manually assessed a subset of the detected information flows concerning their security risk and severity.

\vspace{2mm}

The rest of this paper is structured as follows: Section~\ref{sec:background} describes background information on the technologies used for the conformance analysis; Section~\ref{sec:methodology} presents the methodology of the conducted study; Section~\ref{sec:results} contains the results of the conformance analysis and the risk assessment of information flows; Section~\ref{sec:discussion} discusses the presented results; Section~\ref{sec:related} presents the related work; and Section~\ref{sec:conclusion} concludes this paper.

\section{Background}
\label{sec:background}

In this work, we use CodeQL to perform taint analysis on nodes from the Node-RED framework and conduct a conformance analysis based on the results.
The background required to understand the presented results is introduced in the following.

\subsection{Node-RED}
\label{sub:node_red}

Node-RED is an open source project originally developed by IBM that allows the creation of flow-based IoT applications.
The platform follows the low-code paradigm, i.e., applications can be developed with \textit{nodes}, which are software components that realize specific functionalities and interact with other such nodes via input and output ports.
A multitude of nodes are available that can, for example, connect with hardware sensors and actuators, access online services such as mail servers or social media platforms, or interact with system resources such as files.
The job of the application developer is mainly visual.
The developer selects nodes 
and wires them together into an application (called \textit{flow} in the Node-RED ecosystem) 
that realizes the intended business functionality.
Only in exceptional cases does the developer have to resort to writing JavaScript code themselves, for example, by using the ``function'' node, which follows the function-as-a-service paradigm.

Nodes are made available for use in this process via the \textit{Node-RED library}. 
It contains some official and many third party nodes that are created by independent developers.
The exact number of official nodes can not reliably be determined, but the vast majority of available nodes are developed by open-source contributors.

A node consists of JavaScript/TypeScript files, an HTML file, and an additional JSON file for packaging.
The node's functionality is implemented in standard JavaScript or TypeScript and can interact with the framework via the Node-RED runtime API.
The HTML file contains a number of definitions needed for integration into the Node-RED framework, styling configurations for the visual editor, help texts, and other information.
Most importantly for this work, the HTML file also specifies the number of inputs and outputs of the node.
In the following, we call the JavaScript/TypeScript files defining the node's behavior its \textit{implementation} and the HTML file providing information about it its \textit{specification}.

Nodes can communicate with each other to allow the creation of flows, i.e., to combine the functionality of multiple nodes in a meaningful way.
They can do this by means of messages exchanged via the Node-RED runtime API.
The framework provides methods for sending messages and for registering a listener.
The routing is done by the user, i.e., when one node's output is connected to another node's input in the visual dashboard, messages sent by the upstream node are forwarded to the downstream one.
Although a node can only have one input port, multiple upstream nodes can be connected to it. 
In this case, they can be distinguished in the receiving node via fields in the message payload.

As of April 2020, a request must be submitted to Node-RED to add a node to the official node library.
Before, any package published on the package manager for Node.js (\textit{npm~\footnote{\;\url{https://www.npmjs.com/}}}) that contained a specific keyword was listed in the library.
It is not clear if and how requests are vetted, but the requirement for a request is for technical reasons and not to enforce any regulations.
Given the context of the platform, it is unlikely that extensive checks concerning the security or other properties of the packages are applied, if any.

\subsection{CodeQL}
\label{sub:codeql}

Information flow analysis is used to determine how data traverses a program and how it is changed. 
The typical use case is for the identification of information leaks or other unwanted and unforeseen software behavior.
It can detect \textit{sources} and \textit{sinks} of information, i.e., places where information is created or enters the program (e.g., user-provided input) and places where information leaves the system (e.g., when it is sent to a server).
The analysis then identifies possible connections between sources and sinks, i.e., program behavior that leads to information from a source propagating and reaching a sink.
Following the related literature~\cite{Krohn07_flume_information_flo_control}, we refer to sources and sinks collectively as \textit{endpoints}.

CodeQL is an open-source tool distributed by GitHub that enables information flow analysis and is often used to automate security checks.
It works by transforming the analyzed code into a relational database that can be analyzed via a custom query language.
Issues that are to be detected (such as known vulnerabilities or bugs) are expressed as queries and applied to the database.
The query engine returns any findings of occurrences of the expressed pattern.
Internally, CodeQL models the analyzed program as an abstract syntax tree, a data flow graph, and a control flow graph.
All can be used in a single query, making the approach highly customizable and allowing highly expressive queries.
The representations capture any possible flow of information through the program, including flows through transformation operations as known from taint analysis~\cite{Schwartz10_taint_analysis_all_you_want_to_know} (e.g., when tracking an object \textit{alpha}, after an operation \textit{beta} = \textit{alpha} + \textit{x} taint analysis as opposed to normal data flow analysis would continue to track the object \textit{beta} since it is dependent on \textit{alpha}).

In this paper, we use CodeQL to identify information flows in Node-RED nodes, thus revealing all endpoints of a node over which communication is possible during operation.

\subsection{Conformance Analysis}
\label{sub:conformance_analysis}

It is a known issue that the implementation of software systems tends to deviate from the system envisioned during the design phase over time.
Phenomenons like architectural erosion or concept drift lead to a discrepancy between \textit{what is} and \textit{what should be}.
Conformance analysis describes the process of checking for deviations between any two system artifacts.
Various use-cases have been presented in the literature (e.g., \cite{Cao24_catma}).
In the presented work, we check the conformance between the implementation and specifications of nodes in the Node-RED framework.
The specified inputs and outputs are extracted from the nodes' HTML specifications, and the implemented endpoints are identified via an information flow analysis with CodeQL.

The outcome of a conformance analysis can be one of three cases, \textit{convergence}, \textit{absence}, or \textit{divergence}.
In our analysis, more nuanced outcomes are possible since there are two variables that decide this case, the nodes' inputs and outputs.
For comparability and simplicity, we aggregate them into the standard three cases.
In the context of this work, they are defined as:

\begin{itemize}
    \item \textbf{Convergence:} The specification corresponds to the implementation, i.e., the number of detected sources and sinks matches the number of inputs and outputs in the specification.
    \item \textbf{Divergence:} More sources and/or sinks are detected in the implementation than inputs and/or outputs are specified.
    This case also contains samples where only the inputs or the outputs divert while the other one converges or shows absences.
    \item \textbf{Absence:} Fewer sources and/or sinks are detected in the implementation than inputs and/or outputs are specified.
    This case also contains samples where only the inputs or the outputs show absences while the other one converges.
\end{itemize}

The generalization of, e.g., measurements where only the observed inputs divert from the specified ones while the outputs converge leads to a loss of information.
However, in our analysis of the results we did not gain any further insights when considering the more nuanced results and thus opted for simplicity to increase the accessibility of the results.
The interested reader may refer to our replication package for the complete results~\cite{replication_package}.

The case of convergence is the desired one in any conformance analysis.
In the context of this work, absences are an issue mainly from a software engineering viewpoint.
While they are worth investigating, we focus on divergences in this paper.
Divergences are the cases that are most important from a security point of view.
Additional ``hidden'' information flows are potentially security- and privacy-relevant.

\section{Methodology}
\label{sec:methodology}

\begin{figure}
\centering
\includegraphics[width=\linewidth]{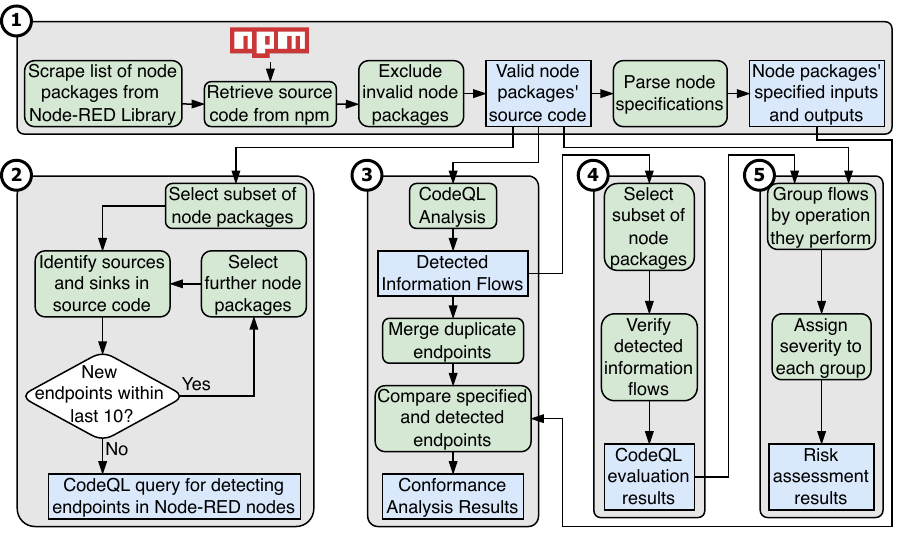}
\caption{Methodology of the conducted study.}
\label{fig:methodology}
\end{figure}

Figure~\ref{fig:methodology} shows the methodology of the conducted work, structured into five steps.
They are presented in detail below.

\subsection{Crawling of the Node-RED library (step \textbf{{\small\Circled{1}}})}
\label{sub:crawling_methodology}

As first step in the performed study, the Node-RED library was crawled to retrieve the source code of all currently listed node packages listed.
A custom script was implemented that scrapes all node package identifiers from the library's official website~\footnote{\url{https://flows.nodered.org/search?type=node}}.
At the time of the analysis, September 10th 2024, 5051 node packages were listed.
The script also retrieves each node package's number of downloads within the last week, which is used to rank their popularity in later steps, the same measure as Node-RED uses.
The analysis script then retrieved the source code of 5032 of the listed node packages from the npm package manager.
The other 19 node packages (0.4\% of the library) had broken download links.
From the downloaded node packages, 133 (2.6\% of the library) were excluded because they did not contain any nodes.
Instead, they contain other content related to Node-RED, such as customizations to change the appearance of the development dashboard or tools that support users in the development.
A further 101 node packages (2.0\% of the library) were removed because they did not contain parsable specifications.
After these steps, the source code and specifications of 4798 valid node packages remained (95.0\% of the library).

\begin{figure}[h]
    \centering
    \begin{minipage}{0.49\textwidth}
        \centering
        \includegraphics[width=\linewidth]{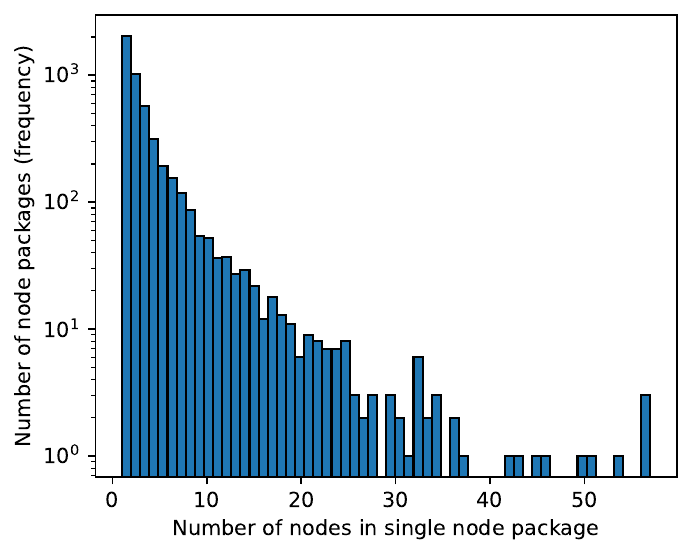}
        \caption{Distribution of nodes per node package. Not showing the ten highest outliers because they skew the graph.}
        \label{fig:nodes_per_package_distribution}
    \end{minipage}
    \hfill
    \begin{minipage}{0.49\textwidth}
        \centering
        \includegraphics[width=\linewidth]{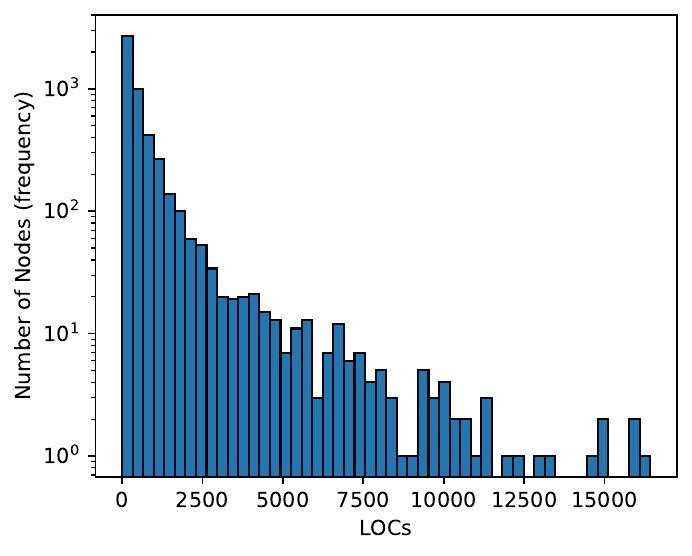}
        \caption{Distribution of the size of nodes (in LOC). Not showing the 100th percentile (51 samples) because it skews the graph.}
        \label{fig:LOCs_per_node}
    \end{minipage}
\end{figure}

The valid node packages contained 17603 individual nodes. 
Figure~\ref{fig:nodes_per_package_distribution} shows the distribution of the number of nodes per node package, excluding the 10 most extreme outliers because they skew the graph.
Four of these outliers are node packages containing more than 100 individual nodes (507, 195, 137, and 134 nodes), the other six range between 61 and 96 nodes.
Most of the valid node packages (2035) contain a single node.

Figure~\ref{fig:LOCs_per_node} shows the distribution of the size of node packages in lines of code (LOC).
The calculation of LOC only considered the .js, .ts, and .html files, i.e., those files that can contain functional code.
The figure does not show the highest percentile of values, since they would skew the graph. 
The largest node package has a size of 148397 LOC. 
Another outlier has 114110 LOC, the remaining 49 values in the 100th percentile are below 100000 LOC.
As shown in the plot, the majority of node packages contains up to 2500 LOC. 

For later use in the conformance analysis, the node packages' specifications were parsed by the analysis script to retrieve the numbers of specified inputs and outputs of each node.
For this, the HTML files in each node package were searched for the parameters \textit{inputs} and \textit{outputs}, considering all different formats of how these parameters are defined throughout the library.

\subsection{Identification of nodes' endpoints (step \textbf{{\small\Circled{2}}})}
\label{sub:endpoint_identification_methodology}

As preliminary for the CodeQL analysis and subsequent conformance analysis, a CodeQL query was required that captures the sources and sinks that CodeQL should consider in its information flow analysis.
The query needed to be created manually, since the endpoints are specific to the Node-RED ecosystem and no such resource could be found in the related literature, Node-RED documentation, or other sources.

To this end, we selected a subset of all valid node packages and manually analyzed their source code to identify sources and sinks of information flows.
The sample size of nodes is chosen with a sample size calculator such that the margin of error is below 10\% with a confidence level of 95\%.
The comparatively high margin of error is tolerated due to the high required manual effort, as indicated by the size of individual node packages shown in Figure~\ref{fig:LOCs_per_node}.
To further strengthen the validity of this step, a saturation criterion was defined: after the initial set of node packages, further ones have to be analyzed until no new sources or sinks are encountered within the last 10 node packages.

Based on the above parameters and the size of the Node-RED library at the time of analysis, there were 97 node packages to be analyzed.
From our experience of the variability of the source code of nodes in the ecosystem, this number is sufficiently high to create a comprehensive CodeQL query.

The samples to be analyzed should cover both the major node packages that are most often used in Node-RED flows and which are often developed and maintained by a team of experienced developers over a long time, as well as less common nodes that are created by independent developers with possibly less rigor.
Therefore, half of the sample were selected from the most popular nodes based on the number of downloads within the last week (this metric is reported and used for ranking by the Node-RED library), and the other half as random node packages from the complete library of valid node packages.

\begin{figure}
    \centering
    \includegraphics[width=0.8\linewidth]{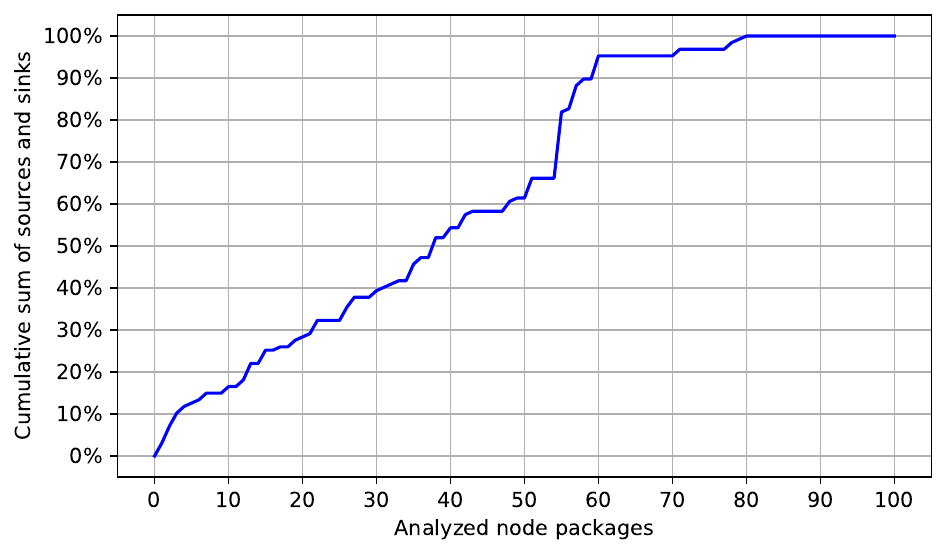}
    \caption{Saturation of manually detected sources and sinks in the analyzed node packages. 100\% $\hat{=}$ 127 endpoints.}
    \label{fig:saturation_plot}
\end{figure}

For the analysis, all JavaScript, TypeScript, and HTML files in the node packages were manually examined.
We identified all methods, variables, and functions in the code that serve as a source or sink (see Section~\ref{sub:codeql}).
Figure~\ref{fig:saturation_plot} visualizes the saturation progression of this process.
It shows the cumulative number of manually detected sources and sinks against the number of analyzed node packages.
As can be seen, 90\% of the overall sources and sinks were identified in the first 60 node packages, and the formulated saturation criterion was met by analyzing the initial 97 node packages already.
Specifically, no new sources or sinks were encountered within the last 20 analyzed node packages.
Thus, no further samples were analyzed afterwards.
We deduce from the process and the saturation plot that the formulated saturation criterion seems applicable and that the manual analysis of 97 node packages looks to be sufficiently exhaustive to base the further conformance analysis on the resulting list of sources and sinks.

The process resulted in the identification of 59 sources and 68 sinks (127 endpoints in total).
They were then translated into a CodeQL query, following the documentation of the query language \textit{QL}~\footnote{\;\url{https://codeql.github.com/docs/}}.
To verify the correctness of the query, we ran it with information flow analysis turned off, that is, so that only the location of detected sources and sinks is reported.
For each of the 127 endpoints, we selected one of the manually identified occurrences of it in the source code and verified that CodeQL had detected this location with the created query.
This process was performed repeatedly during the creation of the query, and possible errors were fixed. 

Although this step required substantial manual effort, it does not have to be repeated if the presented conformance analysis should be repeated in the future, as long as it is reasonable to assume that the source code of node packages is comparable to the current state of the library.

\subsection{Conformance analysis (step \textbf{{\small\Circled{4}}})}
\label{sub:conformance_analysis_methodology}

The conformance analysis of all valid node packages in the Node-RED library of nodes forms the core of this article.
It is based on an information flow analysis performed with CodeQL, which takes as input the source code of the analyzed node packages (as retrieved during step {\small\Circled{1}}) and the CodeQL query file capturing the endpoints that it should detect (as created in step {\small\Circled{2}}).
The analysis is performed on the granularity of node packages instead of individual nodes (see Section~\ref{sub:node_red}).
Although an analysis of individual nodes would allow for a more detailed assessment, information flows detected by CodeQL cannot be reliably assigned to an individual node if it is contained in a node package with other nodes.
While there are clear cases where a single file belongs to each node and hence the association of a detected flow can be done via its location in the source code, there are also many node packages where such a separation is not possible.
For example, if nodes share components such as recurring functionality in custom libraries. 
If an information flow were to be detected in such a file, an additional analysis would have to identify all nodes that access this file and check if this access can trigger the information flow.
The added information gain would not merit the reduced reliability introduced by the additional complexity.
Since the nodes in a node package are created by the same developer or group of developers, we can assume that their level of conformance between implementation and specification should be similar for all nodes in the package.
Therefore, an analysis on the level of node packages is deemed the most precise option.

To perform the CodeQL analysis, each node package was transformed into a CodeQL database, and the created query was applied to each database.
These steps were realized with custom scripts that execute the commands as indicated in the official CodeQL documentation on each node package.
The outcome was an individual CodeQL result file per node package indicating all possible information flows within the node package.

The duplicate endpoints were then merged.
Since CodeQL considers all possible information flows between endpoints, it can detect multiple flows that have the same source and the same sink, for instance, if they differ in a hard-coded input parameter.
In addition, flows can have different sources but the same sink or vice versa. 
Since the argumentation in this article is on the level of individual endpoints, such duplicates had to be resolved to identify only distinct endpoints.
The merging was performed with a custom script that considers the file, line number, and method, function, or variable that form the endpoint to clearly identify the distinct endpoints.
After merging duplicate endpoints, the number of distinct sources and sinks per node package was counted.

Finally, the number of identified sources and sinks was compared against the number of specified inputs and outputs as parsed from the specifications in step {\small\Circled{1}}.
Based on this comparison, the conformance case (convergence, absence, or divergence, see Section~\ref{sub:conformance_analysis}) of each node package was determined and some statistics such as the exact number of additional or missing endpoints were calculated.

\subsection{Evaluation of CodeQL's correctness (step \textbf{{\small\Circled{3}}})}
\label{sub:evaluation_methodology}

To evaluate the accuracy of the analysis pipeline used, we manually checked the correctness of a subset of the information flows that CodeQL detected.
The information flow analysis with CodeQL is the only part of the conducted conformance analysis that is subject to major threats to its validity, due to the difficult nature of such an analysis and the large impact that the quality of the created CodeQL query has on it.
All other steps (retrieving the library of node packages, parsing the specifications, and comparing the specifications against the detected information flows) are simple and robust, and their validity was easily verified during their implementation.

To evaluate CodeQL's accuracy, we selected a subset of all valid node packages following the same process as for the creation of the CodeQL.
We ensured that no node package that had been used to create the query was used for this evaluation of CodeQL as well, since this could thwart the results.
All information flows detected by CodeQL in the thus selected 97 node packages were manually checked for correctness.
Specifically, the locations of the source and sink of each flow reported by CodeQL were examined, and it was checked whether a control flow is possible during the node's execution in which information from the source propagates to the sink.

CodeQL detected 964 information flows across the 97 analyzed node packages.
The manual checking yielded 951 true positives and 13 false positives, resulting in a precision of 0.99.
Considering the chosen margin of error, the precision is seen as sufficiently high to draw conclusions about the Node-RED ecosystem from our conformance analysis based on the evaluated CodeQL information flow analysis.
The false positives detected were caused mainly by ambiguous variable names, e.g., a variable called ``key'' which refers to a key in a dictionary instead of a secret.

Note that while checking for additional information flows not detected by CodeQL would be desirable, such an analysis is extremely labor intensive and error-prone and out of the scope of this work -- compare the sizes of node packages reported in Figure~\ref{fig:LOCs_per_node}.
We hope for the emergence of a dataset from future work that could serve as the ground truth for a study such as the reported one, thus allowing for an evaluation of CodeQL's recall.
Currently, only its precision can be evaluated by us.
Due to this, we cannot make sound statements about absence cases in the conformance analysis and therefore focus on divergence cases in the discussion of the results.

\subsection{Risk assessment (step \textbf{{\small\Circled{5}}})}
\label{sub:risk_assessment_methodology}

Since not all detected divergences necessarily pose a security risk, we assessed the impact of the identified divergent information flows on the security of users.
To this end, the risk associated to a subset of the information flows detected by CodeQL was evaluated.
While the conformance analysis was performed on the level of endpoints, the risk assessment needed to consider complete information flows, to take into account the sensitivity of the transferred data.

The same set of node packages as for the evaluation of CodeQL's correctness (step \Circled{4}) was used for this assessment because these information flows' correctness had been manually verified.
The node packages were classified as different conformance cases.
The selection was not restricted to divergent node packages because it is not possible to distinguish between the information flows within a node that had been considered by its specifications and those that are divergent.
Therefore, information flows from any node package are suitable for this process.

For the risk assessment, the CodeQL results of each sample were analyzed manually, and all detected information flows were classified by the action they perform and the type of data that they send.
In this context, an action can be displaying an error message, writing to a file, sending a message to another node, etc..
The grouping by type of data sent is done via its sensitivity.

The first author performed the classification following an inductive approach, where each assessed information flow was assigned to an existing group if both the performed action and the type of transferred data matched the group, and where a new group was created whenever no suitable one had yet been created.
After assigning each flow to a group, the list was revised by merging groups that could be generalized without loosing specificity about the induced risk, e.g., if the performed action of two groups were different but comparable from the perspective of a risk assessment.
Resulting from this analysis process was a list of groups and a distribution of the analyzed information flows across them.
A severity level was then assigned to each group in a discussion with three authors, indicating the risk introduced by each group of information flows.
The factors influencing the assessment of the severity were: (i) the sensitivity of the transferred data, (ii) the context to which data is exposed (local or outside), and (iii) the possible impact of exposed data on the control flow of the program.
We assumed an attacker model where only the flow development process with the Node-RED framework is confidential, and all other contexts could potentially be observed by an attacker.
We believe this to be reasonable, since the developed programs are deployed on IoT devices that usually run autonomously without supervision and can also be located so that physical access is not controlled.
The assessment followed a worst-case classification in cases of ambiguity where the risk depended on the deployment context.

The classification was performed for all information flows detected by CodeQL, before they are merged in the later stages of the conformance analysis.
We did so to not lose information due to the merging.
If, for example, CodeQL detects two flows for two different information objects from the same input to the same output, these are merged into one in the conformance analysis. 
However, the two information objects can be of different sensitivity and therefore have different security implications when exposed.
Performing the risk assessment before the merging thus provides more accurate results.

\subsection{Threats to validity}
\label{sec:threats}

The conclusions drawn from the results presented in this work are subject to some threats to validity.
We present possible limitations and how we mitigated them in the following.

\subsubsection{Internal}
Most of the results presented are based on the work of one individual researcher, which introduces the risk of researcher bias and errors during the analysis.
We addressed this threat to validity by discussing the critical parts and intermediate results of the analysis process with the other authors.
A second limitation is introduced by the exploratory nature of the conducted study, for which no existing ground truth could be used to validate the correctness of the non-conformance analysis.
As a mitigation, we manually checked a subset of CodeQL's results, which is both the most crucial and most complex part of the analysis.
The important CodeQL query also had to be created by us and was repeatedly checked for its correct capturing of all manually identified sources and sinkd in the analyzed node packages.
The results of the CodeQL evaluation and the iterative creation of the query indicate, that the analysis pipeline functions as intended, as far as we could check it in the given context.

\subsubsection{External}
Since we analyzed the complete population of the investigated domain (i.e., the complete library of Node-RED node packages), the generalization of the presented results is not an issue. 
However, the results should not be generalized to other domains such as other IoT development frameworks without verifying a proper comparability.

\subsubsection{Construct}
Conformance analysis is a suitable tool to investigate compliance between different system artifacts and is commonly used for this purpose.
Therefore, we consider the threat of misinterpreting the very definition of convergence, divergence, and absence.
The scope of our study is analyzing the compliance between the design-level model (HTML specification) and implementation-level model (data flow graph).
In this context (to the best of our knowledge) our interpretations are in-line with the existing literature.
Using the nodes' specifications as comparator for the information flows detected by CodeQL might not be suitable for other analyses because they are developer-created and therefore not reliable, however, this is exactly the mismatch we aimed to investigate. 
Consequently, it is seen as fitting in our case.

Using static analysis might not be suitable for cases where configurations influence how nodes behave at runtime.
Still, it is considered the best tool for the conducted investigation, as such cases should be rare.

\section{Results}
\label{sec:results}

\subsection{Conformance Analysis}
\label{sub:conformance_analysis_results}

The conformance analysis of all valid node packages in the Node-RED library was performed following the methodology presented in Section~\ref{sub:conformance_analysis_methodology}.
The CodeQL information flow analysis of the node packages' source code using the query created in step {\textbf{\Circled{2}}} yielded a total of 58807 detected information flows. 
After merging duplicate endpoints in the flows, 42590 endpoints across all analyzed node packages remained.
Comparing each node package's specified inputs and outputs against the number of identified endpoints resulted in the classification of each node package into one of the three conformance cases convergence, divergence, or absence.
\begin{figure}
    \centering
    \includegraphics[width=0.8\linewidth]{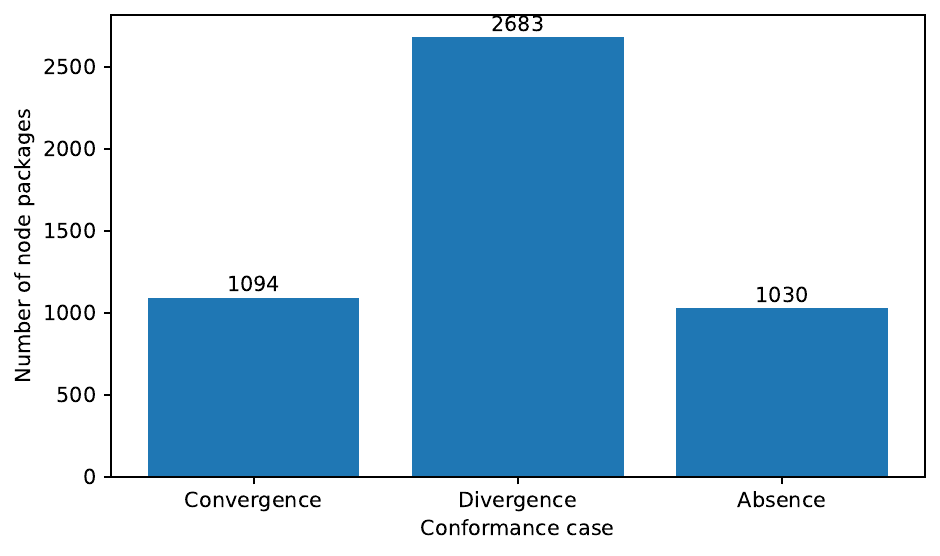}
    \caption{Results of non-conformance analysis, generalized into the three standard cases; n = 4798.}
    \label{fig:non_conformances_generalized}
\end{figure}
Figure~\ref{fig:non_conformances_generalized} shows the results of this classification.
Within the 4798 valid node packages at the time of analysis, 1094 (22.8\%) were classified as convergence cases, 2643 (55\%) as divergence cases, and 1070 (22.2\%) as absence cases. 
In terms of individual nodes contained in the node packages, there are 2240 individual nodes assigned to the convergence case, 10903 to the divergence case, and 4807 to the absence case.
As a ratio of individual nodes per node package, the convergence case has 2.05 individual nodes per node package, the divergence case 4.06, and the absence case 4.67.
Concerning the size of the node packages in lines of code (LOC), the nodes in the convergence case have 270 LOC per node package on average, the nodes in the divergence case 1492 LOC per node package, and the nodes in the absence case 950 LOC per node package.
Comparing the two above measures shows that individual nodes in the convergence case have 132 LOC on average, 367 LOC per node in the divergence case, and 204 LOC per node in the absence case.

\begin{figure}
    \centering
    \includegraphics[width=0.5\linewidth]{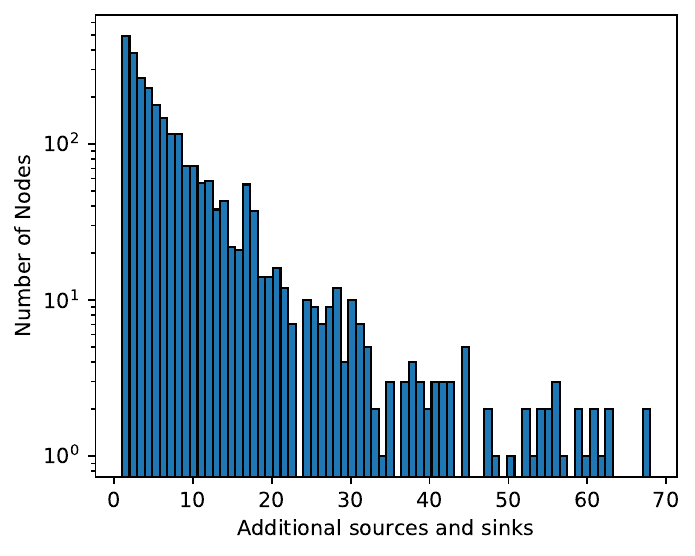}
    \caption{Distribution of the combined number of additional sources and / or sinks per divergent node package as histogram. Not showing the highest percentile because they skew the graph.}
    \label{fig:additional_sources_sinks_histogram}
\end{figure}

Figure~\ref{fig:additional_sources_sinks_histogram} presents a histogram of the number of additional endpoints per node package that is classified as divergent.
It does not show the outliers in the highest percentile for readability purposes. 
The 100th percentile contains 27 values, ten of them are values between 70 and 100, 16 further ones reach up to 194, and the highest value lies at 556 combined additional sources and sinks.
On average, there are 2.9 additional sources and 5.3 additional sinks per node package in the divergent node packages, i.e., 8.2 additional endpoints in sum.
Per individual node, this is an average of 2 additional endpoints per node.

\vspace{2mm}
\noindent
\fcolorbox{black}{green!05}{%
    \parbox{0.983\linewidth}{%
        \textbf{\faExclamationCircle\;RQ1:} Figure~\ref{fig:non_conformances_generalized} shows the distribution of the node packages in the Node-RED library of nodes across the three conformance cases.
        55\% of the node packages are classified as divergent, i.e., containing hidden information flows.
    }%
}%

\subsection{Risk Assessment}
\label{subsec:risk_assessment}

The risk assessment was performed based on the 951 information flows detected by CodeQL within the selected subset of 97 node packages that were manually verified to be correct in step {\small\Circled{4}} of the methodology.
For 12 of the 951, the type of transferred data could not be determined with certainty, and they were therefore not considered in the further assessment, yielding 939 remaining information flows.
By manually analyzing the source code of the node packages and following the methodology presented in Section~\ref{sub:risk_assessment_methodology}, 18 different groups were created to classify the information flows.
\begin{table}
    \small
    \centering
	\caption{Results of the risk assessment. Identified groups in the information flows detected by CodeQL, the number of flows in each group, the percentage of overall flows in the assessment, and the assigned severity of the introduced security risk. Separated by type of the passed information. \textbf{Sev.} = Severity; h/m/l = high/medium/low}
	\label{tbl:risk_assessment_results}
    \begin{tabular}{p{2.1cm}p{7.1cm}p{1.8cm}r}
        \toprule
        \textbf{Information} & \textbf{Group description} & \textbf{Flows} & \textbf{Sev.} \\
        \midrule
        Sensitive  & Display sensitive information in terminal & 37 (3.9\%) & h \\
        information & Display sensitive information in dashboard & 46 (4.9\%) & m \\
         & Log sensitive information & 31 (3.3\%) & h \\
         & Send sensitive information to external server & 31 (3.3\%) & h \\
         & Write sensitive information to file & 3 (0.3\%) & h \\
         & Send sensitive information to framework & 11 (1.2\%) & m \\
        \midrule
        Error  & Log error message & 16 (1.7\%) & h \\
        message & Display error message in dashboard & 110 (11.7\%) & m \\
         & Display error message in terminal & 18 (1.9\%) & h \\
        \midrule
        Input  & Send input message to other node & 323 (34.4\%) & l \\
        message & Log input message & 56 (6.0\%) & h \\
         & Send input message to external hardware device & 2 (0.2\%) & h \\
         & Display input message in dashboard & 175 (18.6\%) & m \\
         & Write input message to file & 6 (0.6\%) & h \\
         & Send input message to external server & 22 (2.3\%) & h\\
         & Display input message in terminal & 28 (3.0\%) & h \\
        \midrule
        Misc. & Misc. low severity & 10 (1.1\%) & l \\
         & Misc. high severity & 14 (1.5\%) & h \\
        \bottomrule
    \end{tabular}
\end{table}
Table~\ref{tbl:risk_assessment_results} presents the groups, the number of information flows that were assigned to each, and the severity assessment associated with each group.

When comparing the different groups, it is apparent that the type of transferred information does not affect the severity, only the context to which it is exposed.
As we realized late during the analysis, each group's type of transferred datacan contain sensitive information.
The first block of groups in Table~\ref{tbl:risk_assessment_results} (with \textit{Sensitive information} in the column \textit{Information}) are groups directly related to sensitive information.
These are information flows that handle entities such as passwords, encryption keys, usernames, etc.
The second block of groups (\textit{Error message} in column \textit{Information}) cover information flows that handle error messages.
While standard error messages of existing libraries and frameworks are usually innocuous, they can also contain information that should not be disclosed.
Especially when they are customized by developers, they often contain sensitive information.
Multiple CWEs refer to this issue (CWE-209 \textit{Generation of Error Message Containing Sensitive Information}~\footnote{https://cwe.mitre.org/data/definitions/209.html} as well as other more specific CWEs) and it has been noted in the literature as well~\cite{Halfond06_error_messages,Smith10_error_messages}.
Finally, the third block of groups in Table~\ref{tbl:risk_assessment_results} (\textit{Input message} in column \textit{Information}) contains groups that handle the input message a node received from other upstream nodes.
These messages also often contain sensitive information required for the node's functionality.
After establishing that all information could potentially be sensitive, the context to which it is exposed remains the only factor to determine the risk imposed by an information flow.
We retain the created grouping nevertheless to give a more nuanced classification.

\begin{table}
    \small
    \centering
	\caption{Results of the risk assessment summarized per severity rating.}
	\label{tbl:risk_assessment_results_summary}
    \begin{tabular}{p{2cm}p{1.5cm}p{1.5cm}}
        \toprule
        \textbf{Severity} & \textbf{Flows} & \textbf{Percentage} \\
        \midrule
        Low & 333 & 35.5\% \\
        Medium & 342 & 36.4\% \\
        High & 264 & 28.1\% \\

        \bottomrule
    \end{tabular}
\end{table}

Table~\ref{tbl:risk_assessment_results_summary} summarizes the results of the risk assessment by the assigned severity rating.
A low severity can be assumed for 35.5\% of the information flows, a medium severity for 36.4\%, and a high severity for 28.1\%.

\vspace{2mm}
\noindent
\fcolorbox{black}{green!05}{%
    \parbox{0.983\linewidth}{%
        \textbf{\faExclamationCircle\;RQ2:} Table~\ref{tbl:risk_assessment_results_summary} provides the answer to RQ2, summarizing the results of the risk assessment of a subset of detected information flows into the severity ratings. 
        A high severity is associated with 28\% of information flows, a medium severity with 36\% of analyzed flows.
    }%
}%

\section{Discussion}
\label{sec:discussion}

\subsection{Divergences in the \node Ecosystem}

The results presented in Section~\ref{sub:conformance_analysis_results} paint a concerning picture of the state of conformance between the implementation and specifications of the nodes in the \node ecosystem.
More than half of the node packages and an even higher proportion of the individual nodes are detected to be divergent, with an average of 8.2 additional endpoints per node package or 2 additional endpoints per individual node.
Many node packages show dozens of additional endpoints in the implementation compared to the specifications.
As the histogram in Figure~\ref{fig:additional_sources_sinks_histogram} shows, most node packages have only one or a few additional endpoints. 
This could suggest that many cases can be explained by minor negligence by the developer instead of a more substantial problem.
The consequences, however, can be the same as for a node package with many additional endpoints.
Overall, the observed numbers show a disconnection between implementation and specifications that make the specifications useless in their current form.

This disconnection can not be explained by differing definitions of inputs and outputs, since a large portion of node packages were also classified as absence cases.
Further, no such definition could be found from Node-RED, therefore, the results at the least show a lack of consensus among the open-source developers.

Since it cannot be determined which endpoints were considered by the developers when creating a node's specifications and therefore, which are the divergent endpoints, discussing the origins of divergences in detail would be speculative. 
However, as an intermediate step of the risk assessment, Table~\ref{tbl:risk_assessment_results} also shows the type of endpoints that are prevalent in the nodes, regardless of the conformance case.
As shown there, most of the identified information flows either display information in the terminal or the development dashboard, or they log information.
Interestingly, these are likely not required for the corresponding nodes' functionalities in the majority of cases.
Instead, they mostly concern the development or maintenance process, both of which are not performed by the typical end-user of the IoT devices but rather the developers of the applications.
This also suggests that a possible mitigation for part of the issue at hand might be to simply remove or disable these parts of the code without breaking their functionality.

The average number of individual nodes per node package and the average number of LOC per node package per conformance case suggest, that smaller, less complex node packages tend to be divergent.
On the other hand, both, the group of absence and divergence node packages show a higher average number of individual nodes per node package.
This could indicate that node packages with a high number of individual nodes have a tendency for non-conformance cases in general.
Intuitively, this observation seems reasonable, given that the increasing complexity and size of the codebase for larger node packages also add mental load for the developer(s) and make it more difficult to keep an overview of the system and all information flows.

\subsection{Risks to Users of the \node Ecosystem}

The additional, ``hidden'' information flows detected in the conformance analysis that are not captured by the specifications could expose sensitive information and allow infiltration of the system.
The risk assessment presented in Section~\ref{subsec:risk_assessment} shows the possible implications of using the analyzed nodes.
A high severity rating is associated with slightly less than a third of the analyzed information flows, a medium and low severity with slightly more than a third each.

As a general observation, the results of the risk assessment indicate a substantial security risk for users of the Node-RED framework.
Without the possibility to gauge the security implications of a specific node, chances are high that users build applications containing nodes with medium or high risk severity.

As shown in Table~\ref{tbl:risk_assessment_results}, roughly one third of all information flows pass messages between nodes, which is the primary intended use and is therefore associated with a low risk severity.
Another large portion of information flows displays information in the terminal or dashboard or logs it.
The corresponding groups of information flow make up 517 of the analyzed information flows, corresponding to 55.1\%.
They are especially interesting because they do not influence the behavior of the developed application, but instead support the development or maintenance process -- in other words: the developer's work.
While logging is important for failure analysis after an incident occurred, and displaying information to the user in real time during operation can be part of the functionality of an application, we estimate that the extent to which these actions are prevalent in the analyzed node packages is not necessary for the node's functionality.
Especially so, when considering that there are dedicated nodes for these use-cases, such as the \textit{debug}-node to display information in the Node-RED dashboard. 
Using a single node or small number of nodes that are made specifically for actions that could have security implications would allow the realization of security mechanisms at a central point, for example, the sanitization of output values or logs.
Instead, our analysis shows that these functionalities are realized with custom implementations in the analyzed nodes.

Other information flows interact with files, external servers, or hardware devices.
It is likely that these are information flows that are fundamental to the implemented functionality of the corresponding nodes, and therefore a security risk that has to be tolerated and mitigated by properly securing them.

\section{Related Work}
\label{sec:related}

\subsubsection{Security Conformance Analysis between Models and Code}
Detecting non-conformances between some model representation of a software system and its implementation in code can help identify security issues at different abstraction levels and during different steps of the software development lifecycle.
For example, Peldszus et al.~\cite{Peldszus19_data_flow_compliance} and Tuma et al.~\cite{Tuma23_security_compliance} presented an approach that supports developers in identifying cases of non-conformance in the security design by automatically detecting mappings between models and implementation, thereby allowing the detection of deviations.
Other approaches have been presented that are aimed to mitigate architectural drift by checking for non-conformances between the designed architecture of a software system and its implementation.
In this context, the problem of architectural drift or erosion describes discrepancies between the designed architecture and the implemented system that emerge over time.
A systematic mapping study performed by Li et al.~\cite{Li22_architecture_erosion} identified multiple approaches that address this issue.
A foundational work for this was presented by Murphy et al.~\cite{Murphy01_software}, which introduces the concept of software reflexion models 
Based on this, Zhong et al.~\cite{zhong23_domico} have proposed an approach that checks UML diagrams for non-conformances at multiple stages during the software development process.
Uzun and Tekinerdogan~\cite{Uzun19_architecture_conformance} proposed a model-based testing approach based on architectural constraints, where violations of these constraints are automatically detected.
Similarly, Marchezan et al.~\cite{Marchezan23_consistency_checks} propose an approach based on consistency rules of different software artifacts by modeling artifacts of different processes in a common model format.
Then, discrepancies between artifacts and consistency rules can automatically be identified.

Contrary to our work, all the above approaches assume a rich model representation of the analyzed software system beyond the specified inputs and outputs to be available for the conformance analysis.

\subsubsection{Code-level Information Flow Security Analysis}
Code-level security analysis omits the creation of a model of the analysed software system and instead works directly on the source code.
A publication by Tabrizi and Pattabiraman~\cite{Tabrizi19_iot_analysis} has shown the effectiveness of such approaches for the security analysis of IoT devices.
The authors compared a design-level security analysis approach based on model-checking against a code-level security analysis approach based on symbolic execution. 
They found, that the code-level analysis outperforms the design-level analysis in terms of accuracy and execution time.

Some approaches have been proposed that support the process of conducting information flow analysis itself.
Manual effort is required to create the information flow analysis specifications, i.e., to initially adjust the available analysis techniques to the specific domain -- in our case, to create the CodeQL query specific to the Node-RED ecosystem.
This issue limits the applicability and reliability of this type of analysis. 
Several approaches have been proposed as a mitigation of this problem, aiming to automate the identification of information flow analysis specifications, e.g., in C\# (Livshits et al.~\cite{Livshits09_merlin_specification_inference}), Python (Chibotaru et al.~\cite{Chibotaru19_taint_specification_learning}), and Android applications (Rasthofer et al.~\cite{Rasthofer14_android_sources_sinks}, Clapp et al.~\cite{Clapp15_mining_information_flow_specifications}).
For JavaScript, Dutta et al.~\cite{dutta2021inspectjs} have proposed a semi-automated approach that only considers the identification of sinks and is therefore not used in this paper.
Staicu et al.~\cite{Staicu20_taint_specifications} proposed a fully automated approach for the detection of both, sources and sinks, however, their approach is dynamic and therefore not feasible for our work.
While none of the above approaches were suitable in the context of this work, future research could include the adaption or extension of an existing approach to improve the creation process of the CodeQL query we used, especially if an approach such as our conformance analysis were to be used to implement automated compliance checks in Node-RED.

\subsubsection{Node-RED and npm}
The package manager npm and different security aspects of it have been investigated before and possible mitigations to increase its security have been proposed.
Since Node-RED relies on npm, some of the more general findings concern it as well.
For example, Zahan et al.~\cite{Zahan22_npm_weak_links} proposed multiple indicators for the detection of malicious packages on npm.
Decan et al.~\cite{Decan18_npm_vulnerabilities} as well as Alfadel et al.~\cite{Alfadel20_npm_vulnerable_dependencies} investigated security vulnerabilities, and Zimmermann et al.~\cite{Zimmermann19_npm_security_threats} investigated security risks in npm packages and their origin, all largely focusing on the issues of vulnerable dependencies.
All three publications reveal wide-spread issues of vulnerable and exploitable packages in npm while also aiming to support developers in the identification of pressing security issues that need to be addressed.
Ferrerira et al.~\cite{Ferreira21_npm_permission_system} have studied packages in npm and identified, that a stricter access control scheme following the principle of least-privilege could strengthen the security of applications built from npm packages, which applies to the Node-RED framework.
They proposed a permission system and enforcement mechanisms that could protect from malicious packages.
It might therefore be a candidate to help address the issue identified in this paper.

Some reseearch has also been conducted on Node-RED specifically.
A suite of multiple tools designed to increase security and privacy for applications built with the Node-RED framework has been presented by Ioannidis et al.\cite{Ioannidis23_securing_the_flow}. 
Specifically, the authors provide techniques and tools to enable encrypted information flows between nodes, to perform code verification at runtime, to enforce access control policies in Node-RED applications, and to monitor Node-RED applications at runtime to detect security incidents.
While these are valuable tools to enforce security in the ecosystem, their does not investigate the extent to which security issues could occur.
Ancona et al.~\cite{Ancona18_runtime_monitoring} proposed an approach for runtime monitoring of Node-RED applications.
They use program traces to verify that API patterns comply with reference specifications in order to prevent unsafe program flows.
Clerrisi et al.~\cite{Clerissi18_testing_nodered} follow the same idea but also capture static system information in their system models.
Although these techniques do not specifically consider security, their dynamic  approaches complement the static analysis approach we followed in our work well, allowing a holistic security analysis if such approaches were to be combined.
An approach proposed by Ahmadpanah et al.~\cite{Ahmadpanah21_securing_nodered} goes beyond monitoring and instead enforces access control of modules and APIs based on pre-defined security policies via allow-lists.
This idea resembles what we envisioned as one of the possible mitigations to the problem identified in this paper.

To the best of our knowledge, no investigation of the state of security of Node-RED nodes has been published before. 
This paper is the first to perform a comprehensive analysis of the source code that Node-RED applications are made from, concerning its information flow security.

\section{Possible Mitigations and Research Outlook}
\label{sec:conclusion}

Insecure applications running on IoT devices can have grave implications for users.
In this paper, we presented the results of a conformance analysis of nodes in the Node-RED library of nodes.
Comparing the number of specified inputs and outputs against the number of endpoints detected via an information flow analysis with CodeQL, we identified a ratio of 55\% divergent node packages within the complete library of 4798 node packages at the time of analysis.
These divergent node packages exhibit more possible information flow endpoints than are captured by their specifications, thus enabling ``hidden'' information flows.
Such cases could potentially be exploited by attackers to obtain access to sensitive information that should be restricted from them.

A risk assessment of a subset of the information flows showed, that 28\% of them are associated with a high severity rating, imposing a substantial security risk to users of the Node-RED ecosystem.

In a low-code environment such as Node-RED, the impact of security issues is heightened because the user is likely not capable of addressing or even realizing such problems.
We see two main directions to mitigate the identified issue: (1) improve the security of nodes and (2) improve the information provided about nodes and the information available to developers.

Improving the security of nodes could be managed centrally by the framework itself, be left to the open-source developers, or realized in a mixture of these two.
On the framework side, each contributed node could have to undergo a security analysis before being published.
This analysis could, for example, resemble the information flow analysis we have conducted in this paper.
If such a process is in place, it could also be extended with further security checks, such as the execution of a number of industry-standard static application security testing tools (SAST).
The responsibility to perform this process could also be moved to the developers, who would have to run a standardized information flow analysis or other security checks and provide the results of it when submitting new or changed code to the framework.
With such a mechanism in place, contributions could be rejected if they do not meet certain requirements, such as the realization of information flows only with whitelisted functionalities.

Another technical solution could be to adjust the framework to be more restrictive in regard to information flows. 
Nodes could be encapsulated more strictly and data only allowed to enter or exit the nodes via framework-specific channels. 
Such a solution would break the functionality of many nodes that are not adjusted to the new framework version, but it would greatly improve the security of developed applications.
Compatibility issues could be mitigated by introducing the new version over an extended period of time in which existing nodes would have to be migrated, a standard practice for such updates.

Looking in the second direction to address the identified issue, a security analysis process could be enforced as described above, but the vetting process would not necessarily enforce any restrictions based on the analysis.
Simply providing the results of the analysis to users would allow them to make an informed decision about whether to use a specific node or refrain from it.
This approach essentially matches the current idea of providing specifications for contributed nodes, with the difference that the non-conformances between implementation and specifications would be addressed.

More information should also be made available for developers of nodes.
Currently, the only information about security provided by Node-RED concerns enforcing access control mechanisms to the development dashboard.
No description of what are inputs and outputs of nodes exists.
There are no resources that discuss the security of node package contributions or what their specifications should capture.
We believe it is likely that the lack of available information concerning security in part caused the identified issue in the ecosystem.

\textbf{Research outlook:}
\begin{itemize}[label=\textbullet]
    \item Ecosystems that rely on open-source developers' contributions (such as Node-RED) should provide clear instructions on how security-critical functionalities should be implemented and documented.
    Ambiguities can lead to issues when interpretations differ.
    \item Such ecosystems should consider moving away from complete freedom for open-source developers and implementing an automated security analysis process for new contributions.
    For example, an analysis pipeline such as the one presented in this paper could be used to objectively determine the numbers of inputs and outputs of a node.
    The results of the analysis can either be used to allow a more informed decision for the user on whether to use a specific node, or even to decide whether the code should be accepted as part of the framework or rejected.
    \item The issue of hidden information flows could also be addressed on a technical level in the framework, e.g., by enforcing a stricter encapsulation of nodes.
    In such a scenario, the framework could provide functionality for all necessary types of data transfer and disallow any other communication channels.
    These official nodes would become the only code that needs to be thoroughly secured with suitable security mechanisms.
\end{itemize}

\bibliographystyle{splncs04}
\bibliography{NodeRED_bib}

\end{document}